\documentclass{article}

\usepackage{arxiv}

\usepackage[utf8]{inputenc} 
\usepackage[T1]{fontenc}    
\usepackage{hyperref}       
\usepackage{url}            
\usepackage{booktabs}       
\usepackage{amsfonts}       
\usepackage{nicefrac}       
\usepackage{microtype}      
\usepackage{lipsum}
\usepackage{graphicx}
\usepackage{subfig}
\usepackage{algorithmicx}
\usepackage{amsmath}
\usepackage{amsfonts}
\usepackage{array} 

\usepackage{algorithm}

\hypersetup{
    colorlinks=true,
    linkcolor=blue,
    filecolor=blue,      
    urlcolor=blue,
    citecolor=blue
}

\usepackage[noend]{algpseudocode}
\usepackage{caption}

\algnewcommand\algorithmicforeach{\textbf{for each}}
\algdef{S}[FOR]{ForEach}[1]{\algorithmicforeach\ #1\ \algorithmicdo}

\graphicspath{ {./images/} }

\title{High resolution microprice estimates from limit orderbook data using hyperdimensional vector Tsetlin Machines}

\author{
 Christian D. Blakely
\\
  Centre for Artificial Intelligence Research \\
  University of Agder\\
  Grimstad, Norway \\
  \texttt{christian.blakely@uia.no} \\
}

\begin{document}
\maketitle
\begin{abstract}
We propose an error-correcting model for the microprice, a high-frequency estimator of
future prices given higher order information of imbalances in the orderbook. The model takes into account a current microprice estimate given the spread and best bid to ask imbalance, and adjusts the microprice based on recent dynamics of higher price rank imbalances. We introduce a computationally fast estimator using a recently proposed hyperdimensional vector Tsetlin machine framework and demonstrate empirically that this estimator can provide a robust estimate of future prices in the orderbook.    
\end{abstract}


\section{Introduction}
High-frequency trading involves algorithms that can react to market data faster than humans, and these algorithms often rely on signals from the limit orderbook (a list of buy and sell orders) in order to determine market making or market taking strategies. Market making strategies, namely strategies that offer liquidity for market participants by using resting limit orders on the bid and ask side of the orderbook typically need to rely on models which indicate in microsecond or even nanosecond speed where the direction of liquidity will be going. Efficient strategies need to be good at typically two things: 1) detecting market spoofing (false signals) and informed buyers/sellers (strong adversarial directions), and 2) future price estimation in order to maximize profits from their liquidity provision. Market making profits are largely driven by crossing spreads between the bid and ask differentials in quick periods of time, typically in microsecond to millisecond resolutions. Providing liquidity for market taking is most efficient in terms of profits when there is minimal amount of false readings of the orderbook that drive supply and demand signals. For this reason accurate and fast microprice algorithms should reflect information not only at the best bid and ask (most liquid supply and demand price information), but also from other higher depth supply and demand stemming from other price ranks.

\subsection{Contributions and paper organization}

In this paper, we develop an extension of the microprice estimator from Stoikov in \cite{stoikov} that includes an additional adaptive component that corrects the microprice based on information from changes in these higher order supply and demand signals. Our contribution to the market microstructure literature is deriving fast higher order supply and demand features that can be structured into a Tsetlin machine model for fast correction of microprice estimates.

In the first part of the paper we give a brief overview of the microprice estimates derived in \cite{stoikov}. We then formulate  additional information that is needed for further correcting the microprice estimation by deriving features from the limit orderbook and show how we encode them into our model as input.  These features come from higher price ranks in the orderbook, but as they can change rapidly at the microsecond level, these features need to be extracted and evaluated in a fast enough manner to be still relevant during the next event of the orderbook. 

A hyperdimensional vector Tsetlin machine framework from \cite{blakely2024hyperdimensionalvectortsetlinmachines} inspired by \cite{halenka2024exploringeffectshyperdimensionalvectors} is then reviewed followed by an in-depth description of the architecture we will use to compute the microprice corrections. We explore numerically the approach by using orderbook data from level 3 orderbook data provided by Databento and demonstrate empirically that the error correction provides a better forecast of the future price at high-frequency resolutions. We also give estimates on speed considerations, as one the strengths of the approach using Tsetlin machines is the considerably fast evaluation times, which would be important if actually deployed in a high-frequency trading pipeline. To conclude the paper, we then give remarks on future work in the direction of possible FPGA implementations as given in \cite{9088251} and online microprice learning using \cite{prescott2023fpgaarchitectureonlinelearning}.

\section{Review of microprice estimation}
\label{sec:microprice}

One of the key interests in market microstructure studies is extracting features from market events in a limit orderbook and deriving information relevant for risk modeling, price forecasting, volatility forecasting, and many other applications.  The challenge with orderbook data is that it is typically very noisy, with little to no signal for making any kind of robust signal or price derivations. For future price discovery in events that take place in the microsecond to millisecond time frame, adjusting orders for market making purposes is a risk optimization procedure. Thus to know what will happen to the price at the next state of the orderbook is critical to minimize risk. Classically, to understand what the "true" future price would be given the current best bid and ask was to formulate a weighted price based on the current spread and the imbalance in orders at the best bid and ask (averaging immediate supply and demand information). However, this is rarely a true reflection of the next price as order sizes tend to be amended or even canceled completely due to reasons such as market spoofing or informed market taking.  Stoikov in \cite{stoikov} derived a much better approach than the weighted mid price called the microprice which combines information from both the spread size and the aggregate order sizes and was shown to be a more robust estimator. \cite{stoikov} claims the mid-price and weighted mid-price are common in finance, but they have several drawbacks, including their high auto-correlation and lack of theoretical justification as true estimators of asset value.

Built by adjusting the mid-price using a function of the order book imbalance and spread, the microprice was shown to be a better predictor of short-term price moves than the traditional mid-price or weighted mid-price. It is constructed as the limit of expected future mid-prices, taking into account the top of the book orderbook state variables imbalance and spread and is computed using a recursive method from historical top of the book data. We give a brief summary of the computation of the microprice which will be used later when introducing a method of adjusting the microprice given higher price rank details from the orderbook.

The mid-price \( M \) is the average of the best bid price \( P_b \) and the best ask price \( P_a \) 
$M = \frac{P_b + P_a}{2}$ with the weighted mid-price \( W \) is a function of the order book imbalance \( I \) is given as:
$W = I P_a + (1 - I) P_b$
where the imbalance \( I \) is calculated as $I = \frac{Q_b}{Q_b + Q_a}$,
and \( Q_b \) is the volume at the best bid and \( Q_a \) is the volume at the best ask.
Using this, the microprice \( P_{\text{micro}} \) is defined as the mid-price plus an adjustment term based on the imbalance \( I \) and the spread \( S \):
\begin{equation}
P_{\text{micro}} = M + g(I, S)
\end{equation}
where \( g(I, S) \) is a function that adjusts the mid-price based on the current order book state (imbalance and spread).
The prediction of the \( i \)-th mid-price, denoted \( P_i \), is given by:
\begin{equation}
P_i(t) = M_t + \sum_{k=1}^{i} g_k(I_t, S_t)
\end{equation}
where \( g_1(I, S) \) is the first-order adjustment $g_1(I, S) = \mathbb{E}[M_{\tau_1} - M_t | I_t = I, S_t = S]$, with
higher-order adjustments \( g_{i+1}(I, S) \) computed recursively as:
\begin{equation}
g_{i+1}(I, S) = \mathbb{E}[g_i(I_{\tau_1}, S_{\tau_1}) | I_t = I, S_t = S]
\end{equation}
The microprice converges if the following condition holds:
\begin{equation}\label{condition}
B^* G_1 = 0,
\end{equation}
where \( B^* \) is the limit of a matrix \( B \) representing the transition probabilities in a Markov process and \( G_1 \) is the first-order adjustment matrix. The estimation procedure used to compute the microprice adjustments follows these steps:

\begin{enumerate}
    \item \textbf{Symmetrize the data.} For every observation \( (I_t, S_t, I_{t+1}, S_{t+1}, \Delta M) \), generate a corresponding symmetric observation:
    \[
    (1 - I_t, S_t, 1 - I_{t+1}, S_{t+1}, -\Delta M).
    \]
    This symmetrization ensures that the condition \ref{condition} is satisfied, which guarantees convergence of the microprice.

    \item \textbf{Estimate the transition matrices} \( Q \), \( T \), and \( R \). These matrices represent:
    \begin{itemize}
        \item \( Q_{xy} \): Probability of a state transition from \( x \) to \( y \) without a change in the mid-price.
        \item \( T_{xy} \): Probability of a state transition from \( x \) to \( y \) when the mid-price changes.
        \item \( R_{xk} \): Probability that the mid-price changes by \( k \) when the state is \( x \).
    \end{itemize}
    
    \item \textbf{Compute the first-order adjustment matrix} \( G_1 \):
    \[
    G_1 = (1 - Q)^{-1} R K,
    \]
    where \( K \) is the vector of possible mid-price changes.

    \item \textbf{Compute the transition matrix} \( B \):
    \[
    B = (1 - Q)^{-1} T.
    \]

    \item \textbf{Compute the microprice adjustment} \( G^* \) as the sum of the recursive microprice adjustments:
    \begin{equation}\label{microprice}
    G^* = P_{\text{micro}} - M = G_1 + \sum_{i=1}^{\infty} B^i G_1.
    \end{equation}
\end{enumerate}
    
In practice, this sum converges quickly and does not require too many iterations. Of course, this requires enough historical data to make sense from a computational perspective. In our numerical examples, we give some ranges of in-sample data from our empirical studies that make sense for small tick securities. 

\section{Orderbook data processing}

We now show how to extract and process the additional features from the orderbook that we will use in our microprice adjustment algorithm. Let us consider an orderbook with price ranks up to level \( L \) on both the bid and ask sides. Define:
\begin{itemize}
    \item \( A_i \) as the volume at price rank \( i \) on the ask side, for \( i = 1, 2, \dots, L \), where \( i = 1 \) represents the best ask, \( i = 2 \) represents the next price level above the best ask, and so on.
    \item \( B_i \) as the volume at price rank \( i \) on the bid side, for \( i = 1, 2, \dots, L \), where \( i = 1 \) represents the best bid, \( i = 2 \) represents the next price level below the best bid, and so on.
\end{itemize}

The total volume across all price ranks on both the bid and ask sides is:
\[
V_{\text{total}} = \sum_{i=1}^{L} A_i + \sum_{i=1}^{L} B_i
\]
For each price rank, the proportion of the total volume that exists at price rank \( i \) on the ask side, denoted \( P_{A,i} \), is given by:
\[
P_{A,i} = \frac{A_i}{V_{\text{total}}}, \quad \text{for } i = 1, 2, \dots, L
\]
Similarly, the proportion of the total volume that exists at price rank \( i \) on the bid side, denoted \( P_{B,i} \), is given by:
\[
P_{B,i} = \frac{B_i}{V_{\text{total}}}, \quad \text{for } i = 1, 2, \dots, L
\]

As an example, if we have the following volumes:
\begin{itemize}
    \item On the ask side: \( A_3 = 10 \), \( A_2 = 10 \), and \( A_1 = 20 \),
    \item On the bid side: \( B_3 = 10 \), \( B_2 = 10 \), and \( B_1 = 20 \),
\end{itemize}
then the total volume is $V_{\text{total}} = 10 + 10 + 20 + 20 + 10 + 10 = 80$ and 
the percentage of total volume at each price rank on the ask side is:
\[
P_{A,3} = .125, \quad P_{A,2} = .125, \quad P_{A,1} = .25
\]
And the percentage of total volume at each price rank on the bid side is:
\[
P_{B,3} = .125, \quad P_{B,2} = .125, \quad P_{B,1} = .25
\]

\begin{figure}[!ht]
\centering
\includegraphics[width=.7\textwidth]{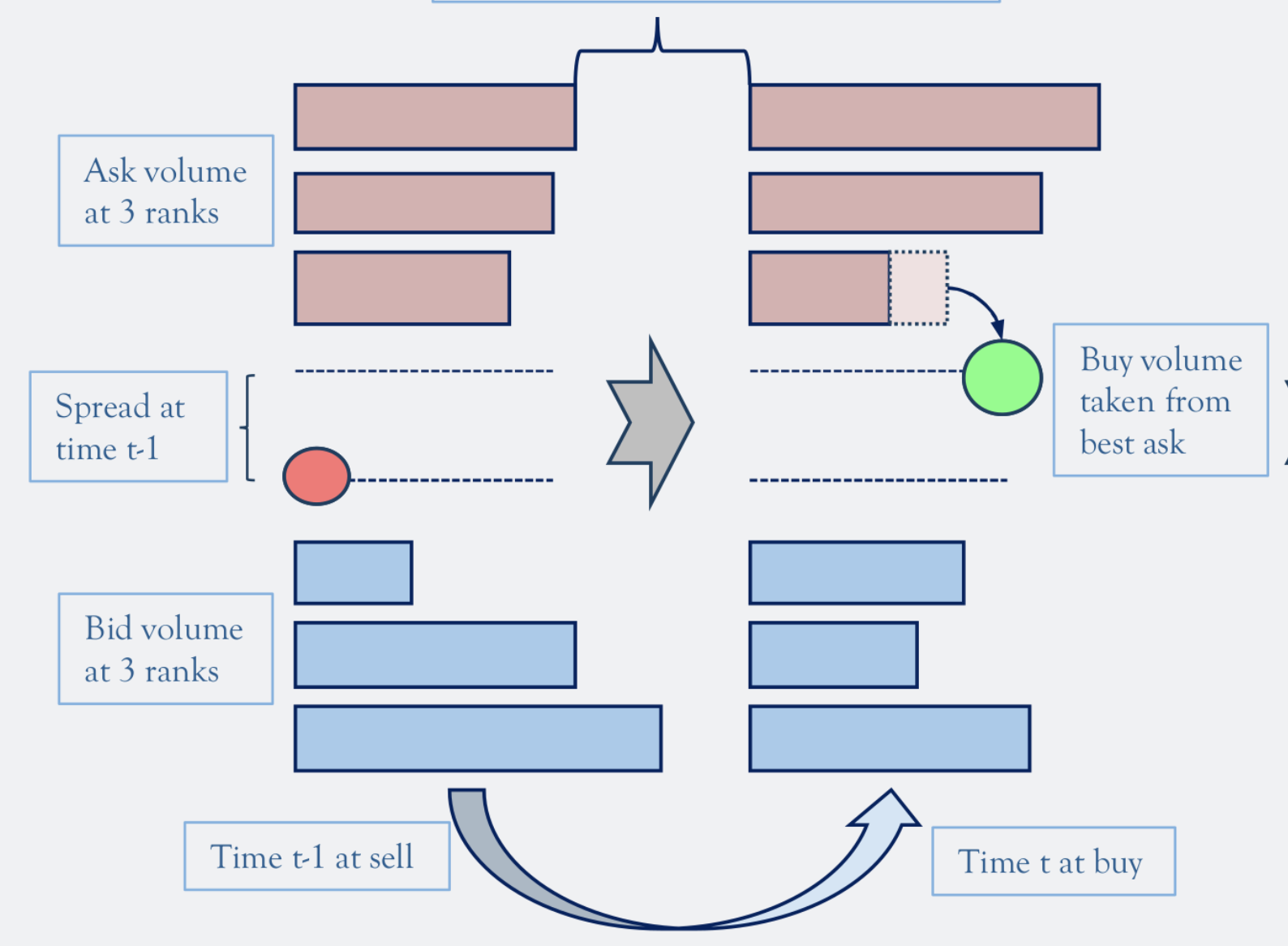}
\caption{The bid and ask side of the orderbook with a buy matching a sell order as an event.}
\label{ref:orderbook}
\end{figure}

For every new event in the orderbook up to price rank $L$, the percentages are adapted according to the event volume inserted, amended, or canceled.  Figure \ref{ref:orderbook} shows the different components of the limit orderbook with an event creating a new snapshot of the orderbook. The features are derived at the first state of the orderbook and then in the second state after an order is matched, the orderbook is updated. 
Notice that the information at these price ranks do not concern the price itself, nor the spread between each price rank, we are merely expressing changes in the supply and demand according to the updates in the supply and demand. Updates in the demand, will also affect the supply, and vice-versa. In addition to the information deriving the microprice estimate (spread and imbalance at best bid and ask), and the supply and demand change increments, we also wish to extract information on an latest price changes based on the market matching at the best bid or ask. When there is a match, and volume is liquidated, the price rank percentage volumes will all change. The price will move a certain number of ticks away for the mid price after a certain number of steps $S > 0$. We take into account this information as well for our microprice adjustment. 

We define the feature vector of the order book, \( \mathbf{F} \), to include the bid and ask volume percentages at price ranks up to \( L \), the spread, and the latest price change in ticks, and the microprice. The feature vector at any timestamp $t$ is expressed as follows:

\[
\mathbf{F}_t = \left( P_{A,1}, P_{A,2}, \dots, P_{A,L}, P_{B,1}, P_{B,2}, \dots, P_{B,L}, S, \Delta M_t , G_t^* \right)^\top
\]
where at timestamp $t$:
\begin{itemize}
    \item \( P_{A,i} = \frac{A_i}{V_{\text{total}}} \) represents the percentage of total volume at price rank \( i \) on the ask side, for \( i = 1, 2, \dots, L \),
    \item \( P_{B,i} = \frac{B_i}{V_{\text{total}}} \) represents the percentage of total volume at price rank \( i \) on the bid side, for \( i = 1, 2, \dots, L \),
    \item \( S = P_a - P_b \) is the spread, which is the difference between the best ask price \( P_a \) and the best bid price \( P_b \),
    \item \( \Delta M_t \) is the latest price change at timestamp $t$, measured in ticks 
    \item $G_t^*$ is the microprice estimate at timestamp $t$
\end{itemize}

With this additional information from higher price-ranks, we now need to encode all this information in an efficient manner. To do this, we utilize hyperdimensional vectors which have several key advantages for orderbook updates:
\begin{itemize}
\item \textbf{Robustness}: HVs are resilient to noise and errors. Small perturbations in the vectors do not significantly affect their overall similarity, making HVC robust to the noisy data of orderbooks
  \item \textbf{Scalability}: HVC can easily handle high-dimensional data, and the operations on HVs are computationally efficient.
  \item \textbf{Simplicity}: The operations on HVs, such as bundling, binding, and similarity measurement, are straightforward and can be implemented efficiently.
  \item \textbf{Speed}: HVs can be effectively represented in binary form, making them attractive to FPGA implementations 
\end{itemize}

\subsection{Encoding Orderbook as a sparse binary hyperdimensional vector}
The complete feature vector \( \mathbf{F} \) consists of:
\begin{itemize}
    \item \( L \) components for the ask-side volume percentages \( (P_{A,1}, P_{A,2}, \dots, P_{A,L}) \),
    \item \( L \) components for the bid-side volume percentages \( (P_{B,1}, P_{B,2}, \dots, P_{B,L}) \),
    \item 1 component for the spread \( S \),
    \item 1 component for the latest price change \( \Delta M \) in ticks.
    \item 1 component for the latest microprice estimate
\end{itemize}

Each of the $2L$ percentages can be quantized to $Q$ different values, while the spread and price change are expressed in number of ticks, and bounded by the largest spread during continuous trading. 

We encode each value in the orderbook feature vector as a sparse binary hyperdimensional vector of dimension \( N \). As we saw earlier, the orderbook feature vector \( \mathbf{F} \) contains volume percentages at price ranks, spread in ticks, price change in ticks, and microprice adjustments which, as we will show now, can readily be quantized and represented by hyperdimensional vectors.

Let \( P_{A,i} \) represent the percentage of volume at price rank \( i \) on the ask side, and \( P_{B,i} \) represent the percentage of volume at price rank \( i \) on the bid side. These values fall in the interval \( [0,1] \) and are quantized into \( Q \) levels. Each quantized level corresponds to a unique sparse binary hypervector. Namely, the encoding of the percentage volume \( P_{A,i} \) is given by:
\[
P_{A,i} \rightarrow \mathbf{H}(P_{A,i}) \in \{0,1\}^N
\]
where \( \mathbf{H}(P_{A,i}) \) is the binary hypervector corresponding to the quantized value of \( P_{A,i} \). Similarly, for the bid side:
\[
P_{B,i} \rightarrow \mathbf{H}(P_{B,i}) \in \{0,1\}^N
\]
Thus, for each price rank \( i \), we have a binary hypervector encoding the quantized percentage volume.
We also represent the number of ticks as a hypervector as well. Let \( S_t \) represent the spread at a given timestamp $t$, measured in the number of ticks, thus $S_t \in \mathrm{N}$. Each value of the spread (in ticks) is represented by a random sparse binary hyperdimensional vector.  We build a dictionary mapping integers to a random sparse binary vector. For example, if the spread is \( S = 2 \) ticks, we assign a random sparse hypervector to the integer 2, and define the mapping
\[
S \rightarrow \mathbf{H}(S_t) \in \{0,1\}^N
\]
where \( \mathbf{H}(S_t) \) is a binary hypervector corresponding to the spread of 2 ticks. Similarly, the latest price change at timestamp $t$ \( \Delta M_t \), measured in ticks, is also represented by a random sparse binary hypervector.  Thus each component of the feature vector \( \mathbf{F} \) can now be quantized to their respective hypervectors:
\[
\mathbf{F}_{\text{HD}} = \left( \mathbf{H}(P_{A,1}), \mathbf{H}(P_{A,2}), \dots, \mathbf{H}(P_{A,L}), \mathbf{H}(P_{B,1}), \mathbf{H}(P_{B,2}), \dots, \mathbf{H}(P_{B,L}), \mathbf{H}(S_t), \mathbf{H}(\Delta M_t), \mathbf{H}(G_t^*) \right).
\]
The next step will be to apply various hypervector operations to combine this collection of vectors into one hypervector, encoding all the necessary information. 

To begin, for each price rank \( i \), we bind the hypervector for the volume percentage at rank \( i \) with the permutation vector \( \mathbf{\Pi}_i \). This captures the rank information for the volume at each price level. For the bid side, the volume percentage at price rank \( i \) is bound with the permutation vector \( \mathbf{\Pi}_i \) as follows:
\[
\mathbf{H}_{B,i} = \mathbf{H}(P_{B,i}) \otimes \mathbf{\Pi}_i
\]
where \( \mathbf{H}(P_{B,i}) \) is the hypervector corresponding to the volume percentage at price rank \( i \) on the bid side. Similarly, for the ask side, the volume percentage at price rank \( i \) is bound with the permutation vector \( \mathbf{\Pi}_i \):
\[
\mathbf{H}_{A,i} = \mathbf{H}(P_{A,i}) \otimes \mathbf{\Pi}_i
\]
where \( \mathbf{H}(P_{A,i}) \) is the hypervector corresponding to the volume percentage at price rank \( i \) on the ask side.  To form the final hypervector, we bundle together all the vectors for the volume percentages on both the bid and ask sides, along with the vectors representing other features (spread, price change, and microprice).

The final bundled vector at timestamp $t$ is expressed as:
\[
\mathbf{F_t,}_{\text{HD}} = \left( \bigoplus_{i=1}^{L} \mathbf{H}_{B,i} \right) \oplus \left( \bigoplus_{i=1}^{L} \mathbf{H}_{A,i} \right) \oplus \mathbf{H}(S_t) \oplus \mathbf{H}(\Delta M_t) \oplus \mathbf{H}(G_t^*)
\]
represents the encoded state of the order book, including all volume percentages, spread, price change, and microprice adjustment, with rank information incorporated through the permutation vectors.

Of course, for learning how to adjust the microprice, the encoding of the orderbook feature vector is not the final step. We still need to label the feature vector in order to learn the literal patterns from the Tsetlin machine. We introduce a parameter $s$ which is defined as the number of events in the future to aggregate in order to extract a future price. Here, there are several different choices one can make and depends on the trading application. Here are a few examples for labeling using information bars. We let 
\begin{itemize}
    \item $P_i$ represents the price at trade $i$,
    \item $V_i$ represents the volume at trade $i$,
    \item $s$ represents the number of trades or volume increments, depending on the bar type.
\end{itemize}
Different bars for labels can then be create using the parameter $s$ as
\begin{itemize}
    \item Tick bars are created when a fixed number $N_{\text{ticks}}$ of trades (ticks) have occurred. A new bar is formed when the total number of trades reaches $N_{\text{ticks}}$: $\sum_{i=1}^{s} 1 = N_{\text{ticks}}$
    \item Dollar bars are created when the cumulative dollar value of the traded volume, given by $\sum_{i=1}^{n} P_i \times V_i$, reaches a specified dollar threshold $D_{\text{threshold}}$. A new bar is formed once this dollar value is met: $\sum_{i=1}^{s} P_i \times V_i = D_{\text{threshold}}$
    \item Volume bars are created when the cumulative traded volume $\sum_{i=1}^{n} V_i$ reaches a specified threshold $V_{\text{threshold}}$. A new bar is formed when this total volume is met $\sum_{i=1}^{s} V_i = V_{\text{threshold}}$
\end{itemize}
For the final price, we take the closing price from the information bar, and count how many ticks there are away from the latest microprice estimate.

\begin{figure}[!ht]
\centering
\includegraphics[width=.7\textwidth]{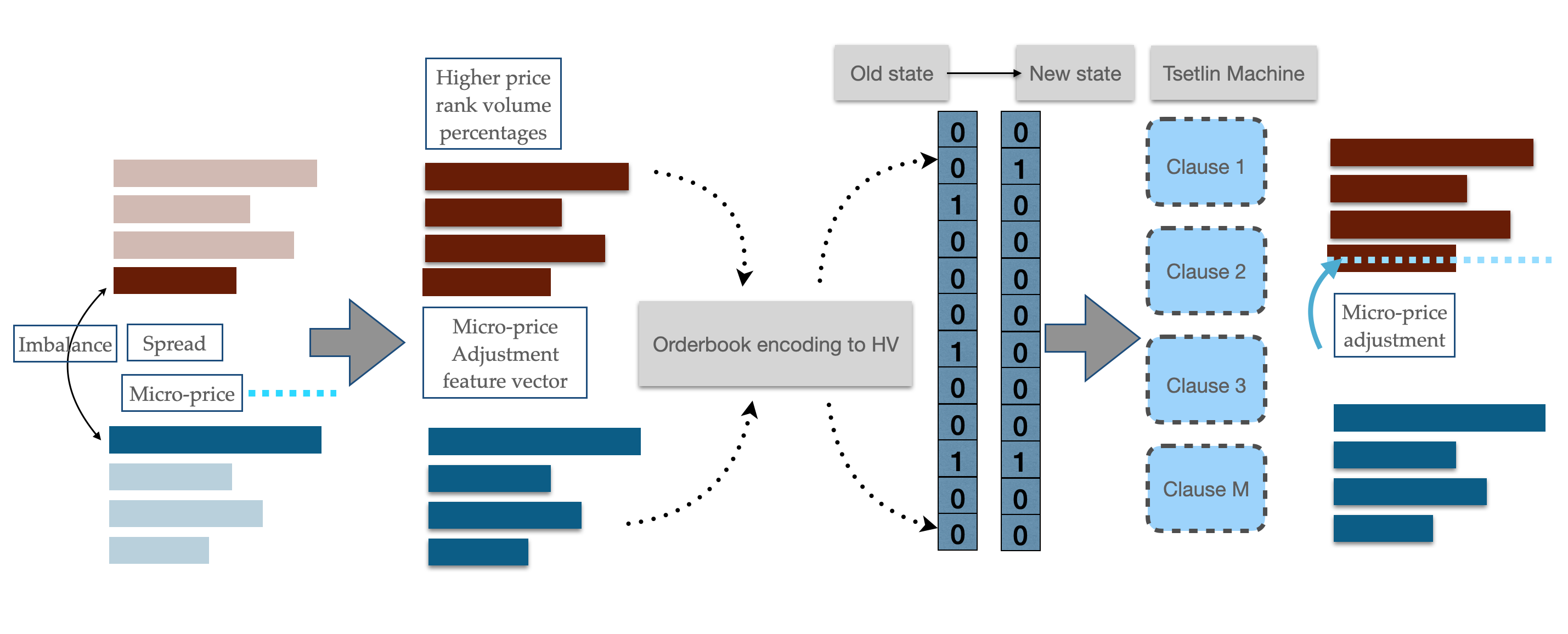}
\caption{Sequence of extracting the feature vector for encoding into a hyperdimensional vector for learning using the Tsetlin machine.}
\label{ref:orderbook2}
\end{figure}

Figure \ref{ref:orderbook2} illustrates the process of extracting the relevant features for learning how to adjust the microprice.  From left to right, the first step constructs the microprice using the current spread and imbalance values from the best bid and best ask, ignoring the higher price rank information. In the second step, the percentage volumes are computed at the higher price ranks along with any price movements in number of ticks. This information is encoded into an HV and ready for input to the TM which will evaluate what type of adjustment to make to the microprice.  

\begin{figure}[!ht]
\centering
\includegraphics[width=.7\textwidth]{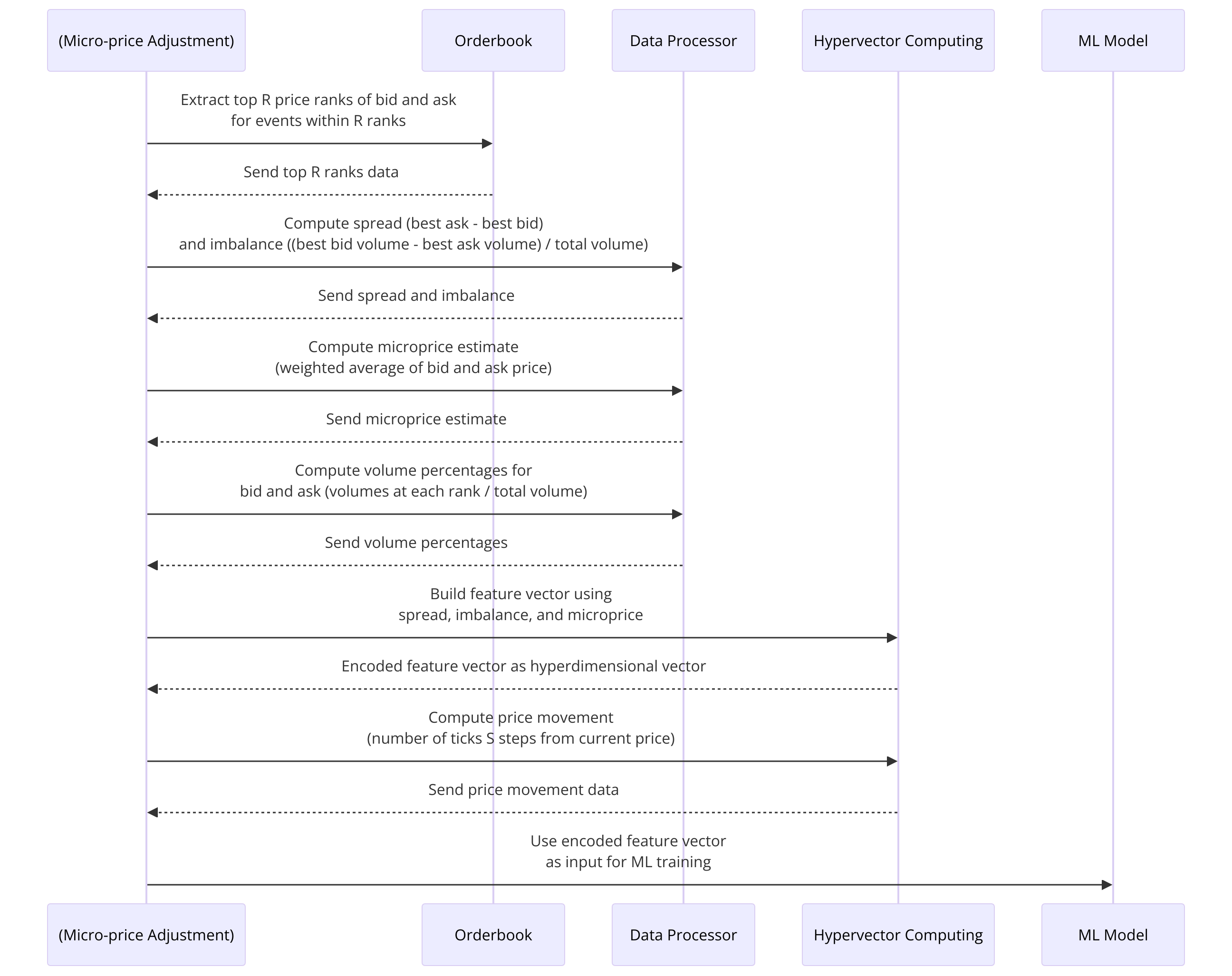}
\caption{Detailed sequence flow of extracting information from raw orderbook data to build the feature vector. Once the feature vector is computed, and the future price in number of ticks from current microprice is computed S steps in the future, the microprice correction is learned.}
\label{ref:orderbook3}
\end{figure}

Figure \ref{ref:orderbook3} shows a sequence diagram of the flow of information in extracting and processing the data necessary for the TM model.

\section{Updating microprice}

\begin{figure}
\centering
\includegraphics[width=.7\textwidth]{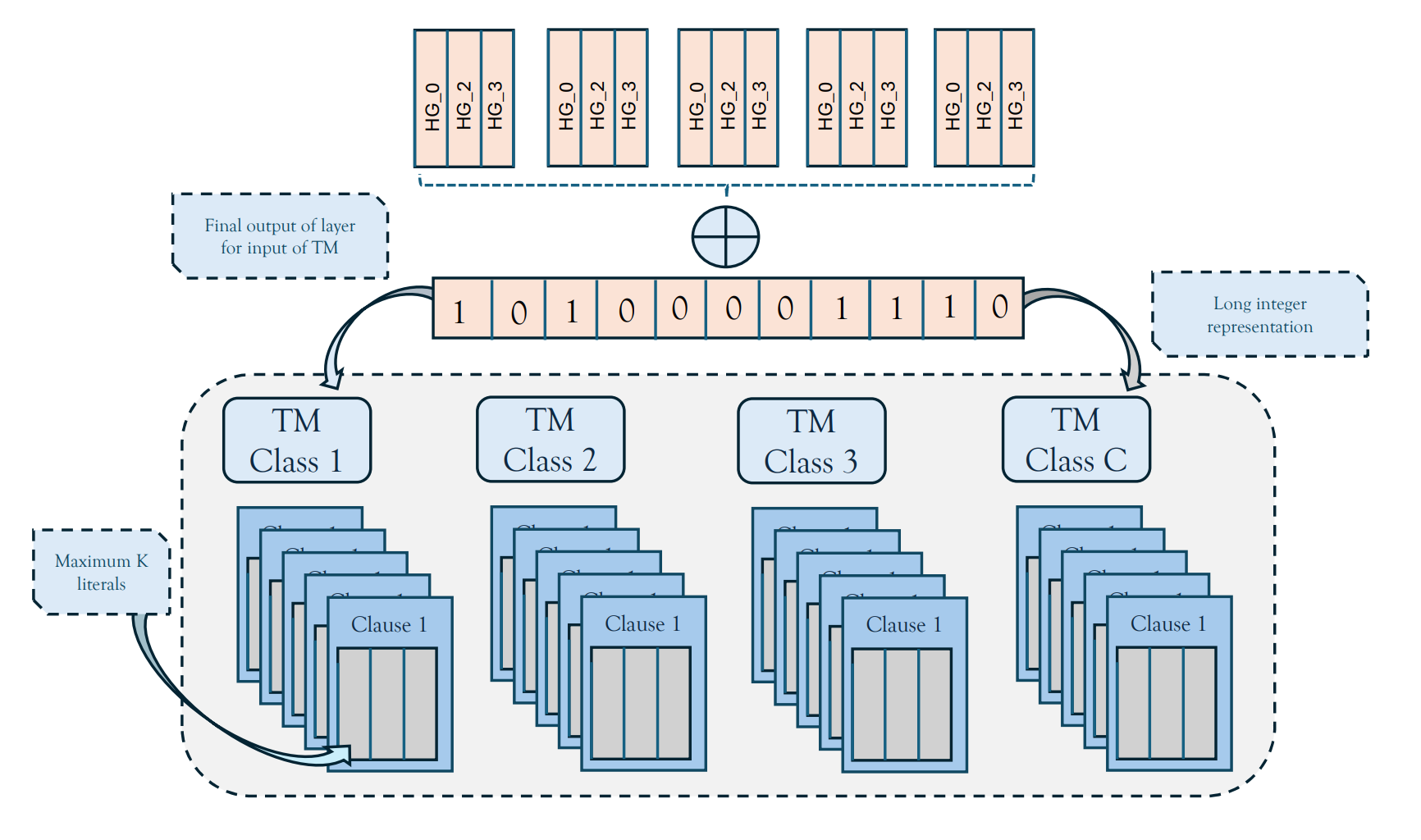}
\caption{Input into the TM is the encoded latest $N$ values of the sequence. This input is used to predict the next value governed by the $Q$-class TM}
\label{ref:tsetlinMachine}
\end{figure}

Updating an estimated microprice to take into account orderbook information in order to adjust for other bid and ask movements comprises of a few computational steps. Assembling the encoded hypervector from the orderbook, binding all the encoded components, and then evaluating the microprice adjustment given the current microprice estimate using the TM framework are given in more detail in Algorithm \ref{adjustementAlgo}.  Figure \ref{ref:tsetlinMachine} depicts the flow of information to the $C$-class TM. 

\begin{algorithm}
\caption{Update Algorithm for event \( E \) at Price Rank \( B_i \) or \( A_i \)}\label{adjustementAlgo}
\begin{algorithmic}[1]
\Require Event \( E \) at price rank \( B_i \) or \( A_i \), where \( 1 \leq i \leq L \)
Ensure Updated microprice, volume percentages, spread, tick change, feature vector, adjusted microprice estimate

\If{$E$ occurs at price rank 1 (i.e., best bid or best ask)}
    \State Update the microprice estimate \( P_{\text{micro}} \) using the microprice formula:
    \[
    P_{\text{micro}} = M + g(I, S)
    \]
    where \( M \) is the mid-price, \( I \) is the imbalance, and \( S \) is the spread.
\EndIf

\For{each price rank \( i \) from 1 to \( L \)}
    \State Recompute the volume percentage \( P_{A,i} \) on the ask side:
    \[
    P_{A,i} = \frac{A_i}{V_{\text{total}}}, \quad \text{where} \quad V_{\text{total}} = \sum_{i=1}^{L} A_i + \sum_{i=1}^{L} B_i
    \]
    \State Recompute the volume percentage \( P_{B,i} \) on the bid side:
    \[
    P_{B,i} = \frac{B_i}{V_{\text{total}}}
    \]
\EndFor

\State \textbf{Step 3: Compute the spread}
The spread \( S \) is computed as:
\[
S = P_a - P_b
\]
where \( P_a \) is the best ask price and \( P_b \) is the best bid price.

\State Compute price change
    \[
    \Delta M = \frac{P_{\text{new}} - P_{\text{old}}}{\text{tick size}}
    \]
\State \textbf{Step 5: Assemble the feature vector}
Assemble the feature vector \( \mathbf{F} \) of order book features, including the newly estimated microprice \( P_{\text{micro}} \):
\[
\mathbf{F} = \left( P_{A,1}, P_{A,2}, \dots, P_{A,L}, P_{B,1}, P_{B,2}, \dots, P_{B,L}, S, \Delta T, G_t^* \right)^\top
\]

\State \textbf{Step 6: Compute the adjusted microprice using hypervector encoding and Tsetlin machine evaluation}
Using hypervector encoding, compute the adjusted microprice estimate \( P_{\text{adj}} \):
\[
P_{\text{adj}} = \text{HyperdimensionalEncoding}(\mathbf{F})
\]

\State \textbf{Step 7: Predict the adjusted microprice using the Tsetlin machine}
Use the Tsetlin machine to predict the adjusted microprice:
\[
P_{\text{Tsetlin}} = \text{TsetlinMachinePrediction}(\mathbf{F})
\]
\end{algorithmic}
\end{algorithm}

Notice in the final step, we assume a fully trained TM in order make the predictions in how to adjust the microprice. The prediction we use here is how many ticks we need to adjust the microprice. With $C$ classes to work with, training the TM using properly labeled data is a crucial element as we need to ensure to extract the most amount of features for each class. Thus we try to keep the number of classes as low as possible. The strategy we employ in our empirical studies is to enforce the following: 
\begin{enumerate}
    \item Class $0 \rightarrow -2$ tick adjustment
    \item Class $1 \rightarrow -1$ tick adjustment
    \item Class $2 \rightarrow$ no adjustment
    \item Class $3 \rightarrow 1$ tick adjustment
    \item Class $4 \rightarrow 2$ tick adjustment
\end{enumerate}
Of course, more (or less) classes can be derived, but the empirical studies on have demonstrated that these are a fairly robust starting point. 

\section{Empirical Studies}

In order to assess any benefits of using the higher order information from the orderbook in order to adjust the microprice, we establish a criterion from which we can build a systematic direct comparison on orderbook information. We will limit our study to equities of different types, namely a smallcap (TEM) and a blue chip (TSLA). Data is provided by Databento. 

Let \( \hat{P}_{\text{micro}, t} \) represent the microprice estimate at time \( t \), and let \( P_{t+N} \) be the true price at time \( t+N \), where \( N \) is the number of price updates into the future. The error at each time step \( t \) is defined as the difference between the microprice estimate and the true price \( N \)-steps into the future:
\[
e_t = \hat{P}_{\text{micro}, t} - P_{t+N}
\]
Using this, the sum of squared errors over a given time window of \( T \) time points is given by:
\[
\sum_{t=1}^{T} e_t^2 = \sum_{t=1}^{T} \left( \hat{P}_{\text{micro}, t} - P_{t+N} \right)^2
\]
where we evaluate this in the \( L_2 \)-norm 
\begin{equation}\label{ref:error}
  \text{Error}_{L_2} = \sqrt{\frac{1}{T} \sum_{t=1}^{T} \left( \hat{P}_{\text{micro}, t} - P_{t+N} \right)^2}  
\end{equation}
This \( L_2 \)-norm error provides a measure of how well the microprice estimates align with the true prices \( N \)-steps into the future and will be used in the empirical studies that follow.
To determine which prices we evaluate the error on, we utilize both the tick bar and volume bar approach to labeling and gathering the training and test data.  

\subsection{Comparison with Microprice}

In comparing the size of the microprice adjustments, it is no surprise to see that the largest adjustment contributions are coming during the early trading sessions after market open. This is when volatility and spreads are at their largest, contributing to the variance in the microprice estimation. The adjustments take into account larger order sizes that get filled at higher bid and ask prices, which in turn will create larger adjustments in the microproce. 

\begin{figure}
    \centering
    \subfloat[Example 1 Microprice Adjustment High Volatility]{{\includegraphics[width=9cm]{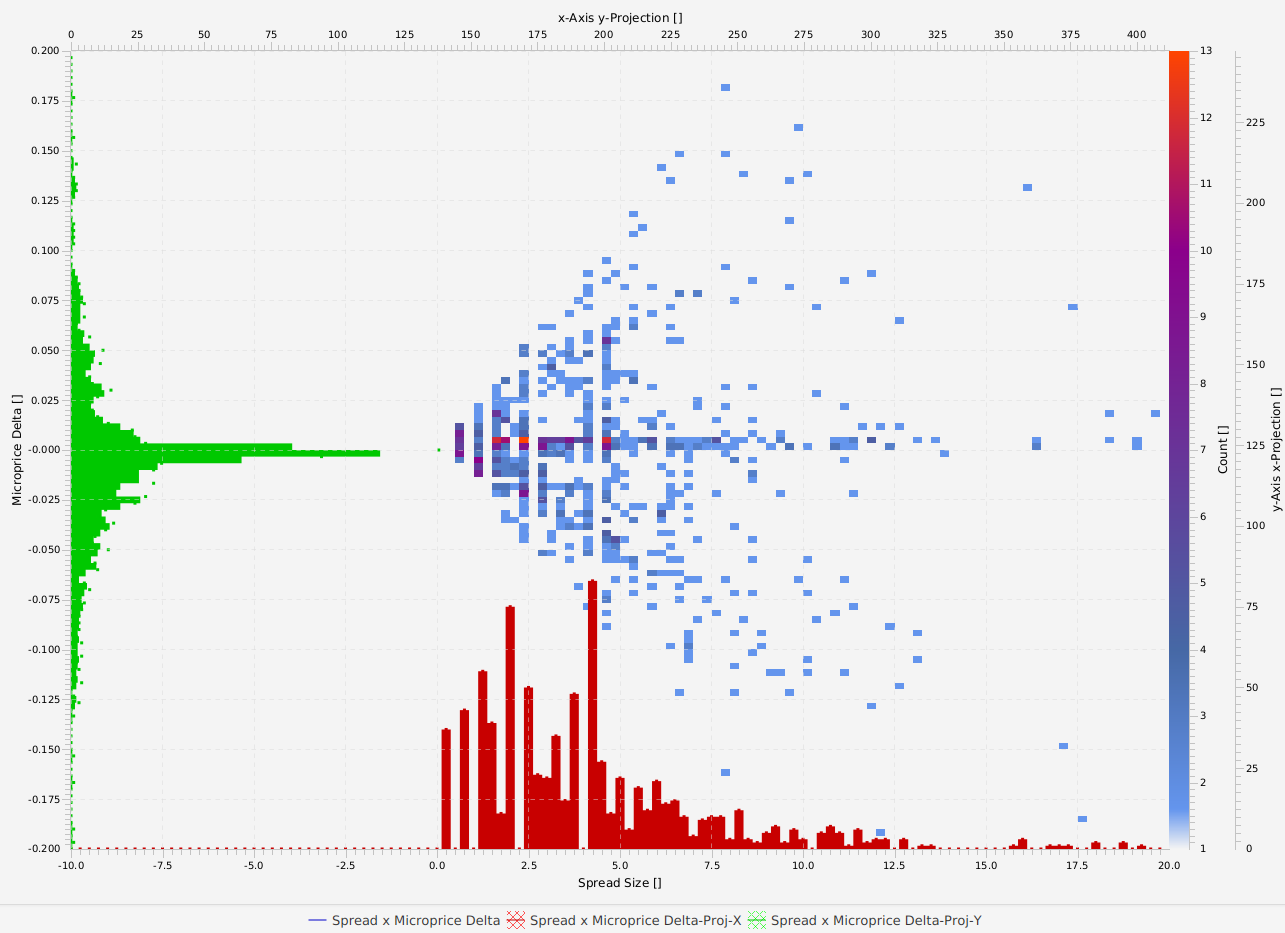} }}%
    \qquad
    \subfloat[Example 2 Microprice Adjustment Lower Volatility]{{\includegraphics[width=9cm]{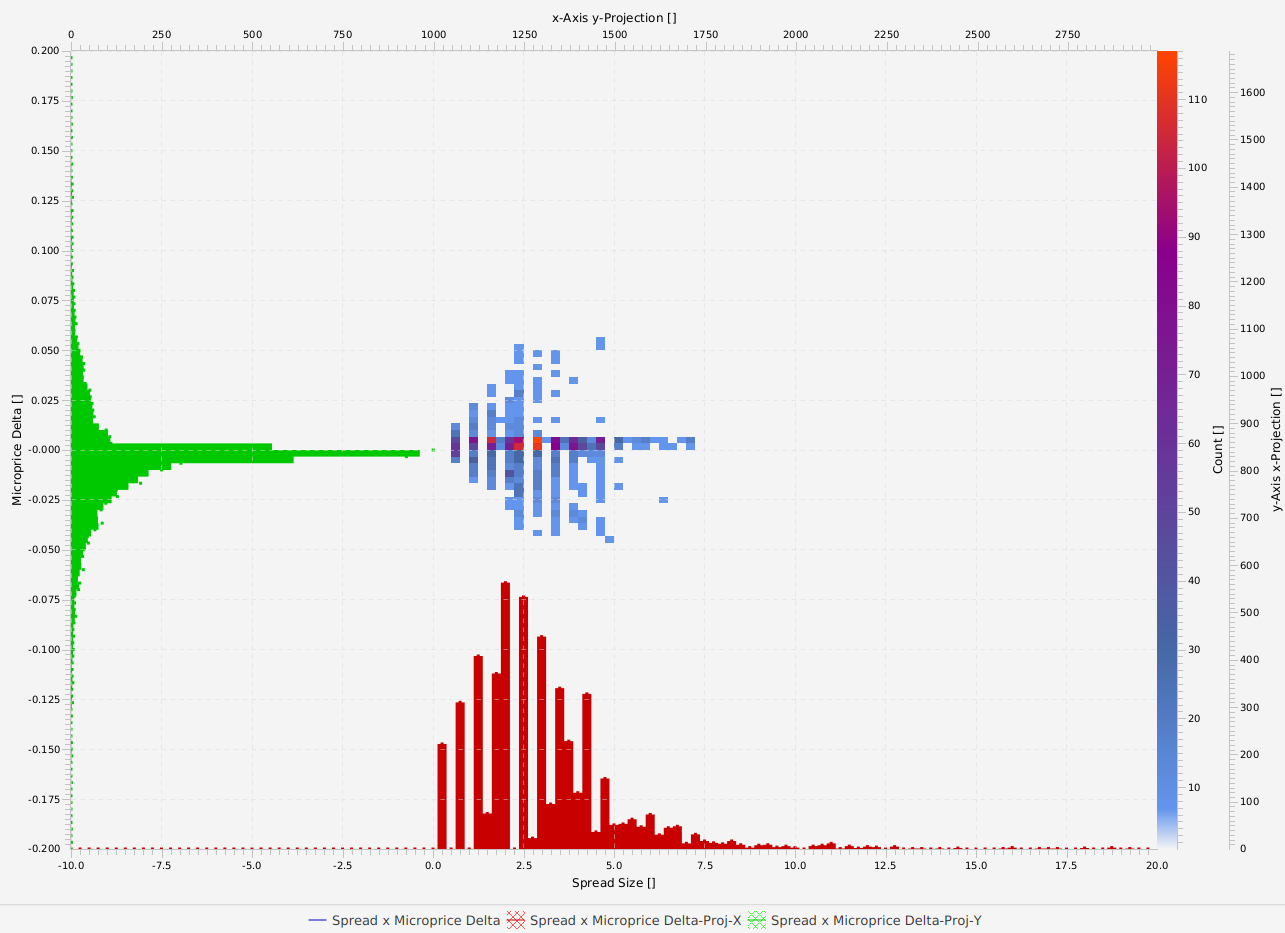} }}%
    \caption{Microprice adjustments for various spreads}%
    \label{fig:spreadMicroprice}%
\end{figure}

In figure \ref{fig:spreadMicroprice} we show an example of microprice adjustments for various spreads over morning market open (example 1) where volatility is higher, and mid day where volatility tends to be lower. 

We also see that the adjusted microprice tends to perform the same as the microprice when spreads are more tight as trading continues during the afternoon trading session as volatility dies down.

\begin{figure}
    \centering
    \subfloat[Imbalance vs Microprice adjustement delta TEM]{{\includegraphics[width=7cm]{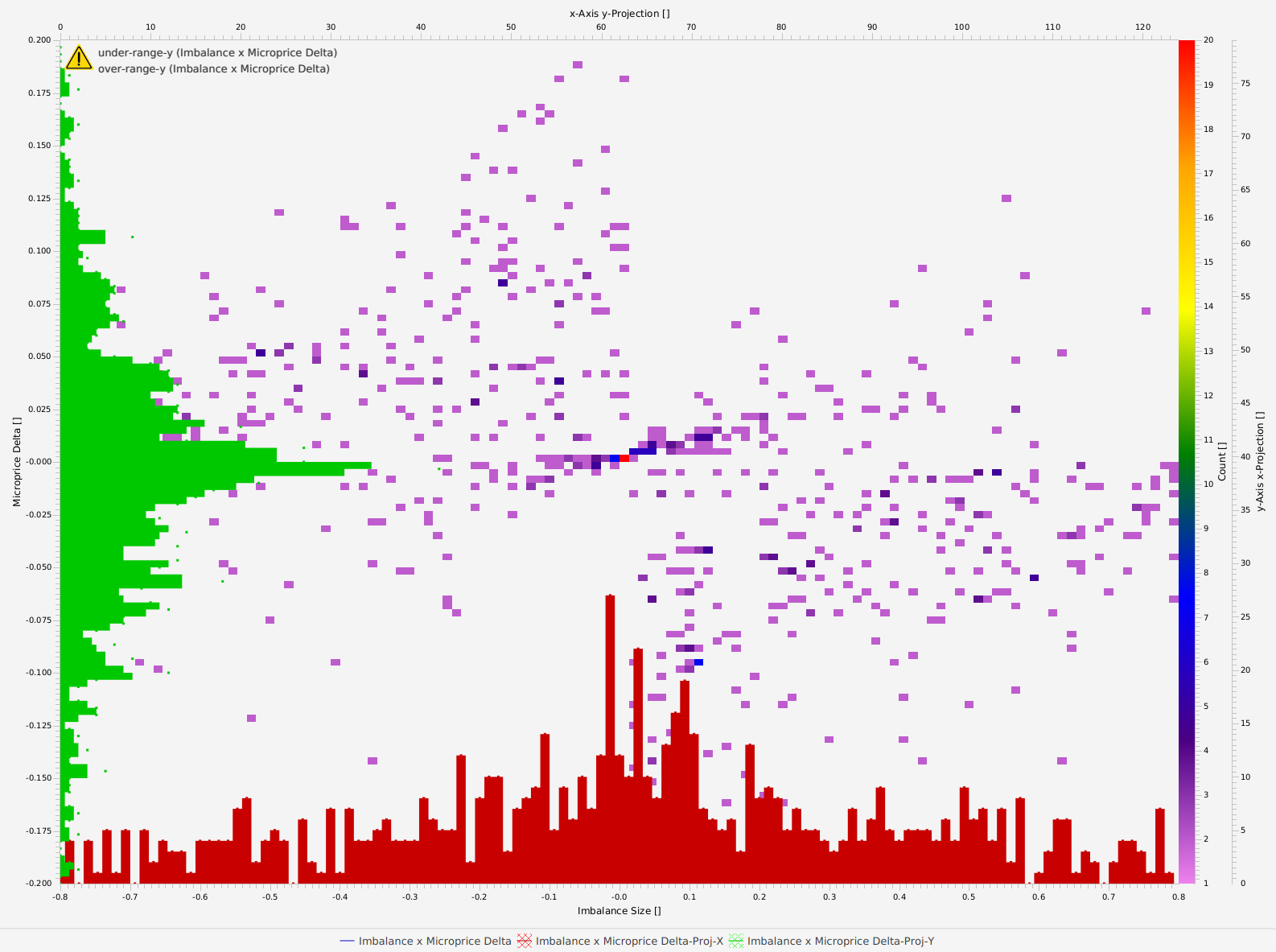} }}%
    \qquad
    \subfloat[Imbalance vs Microprice adjustement delta TSLA]{{\includegraphics[width=7cm]{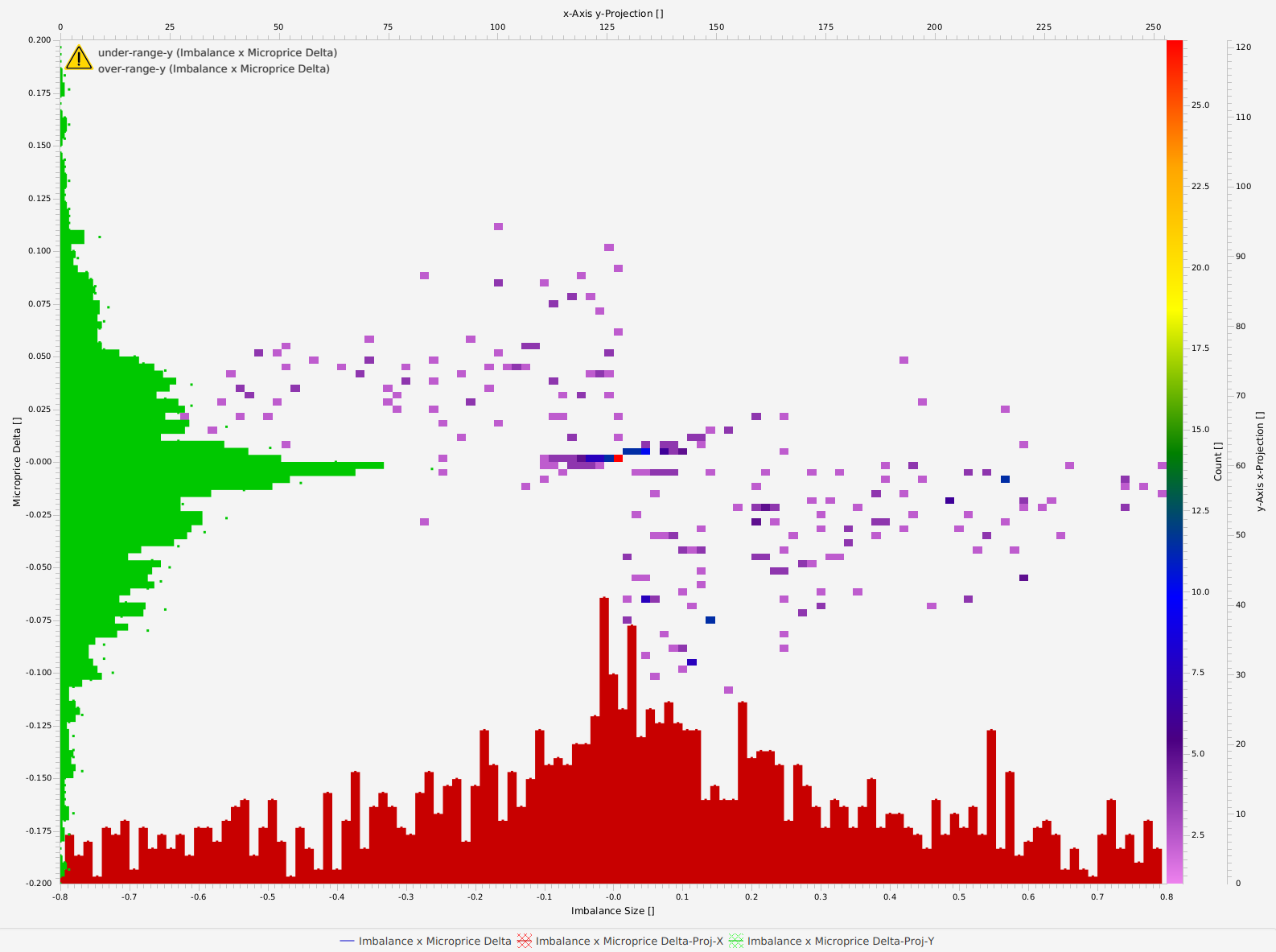} }}%
    \caption{Imbalance vs Microprice delta on SmallCap TEM and BlueChip TSLA}%
    \label{fig:imbalanceMicroprice}%
\end{figure}

Since microprice adjustment is a function of orderbook imbalance as well as spread size, we show the some samples over morning trading over several days on the difference in the microprice adjustment estimate and the latest matched price. We see in \ref{fig:imbalanceMicroprice} the imbalance against microprice delta on TEM how scattered the change is regardless of the imbalance to the ask or the bid side of the orderbook.

\begin{figure}
\centering
\includegraphics[width=.7\textwidth]{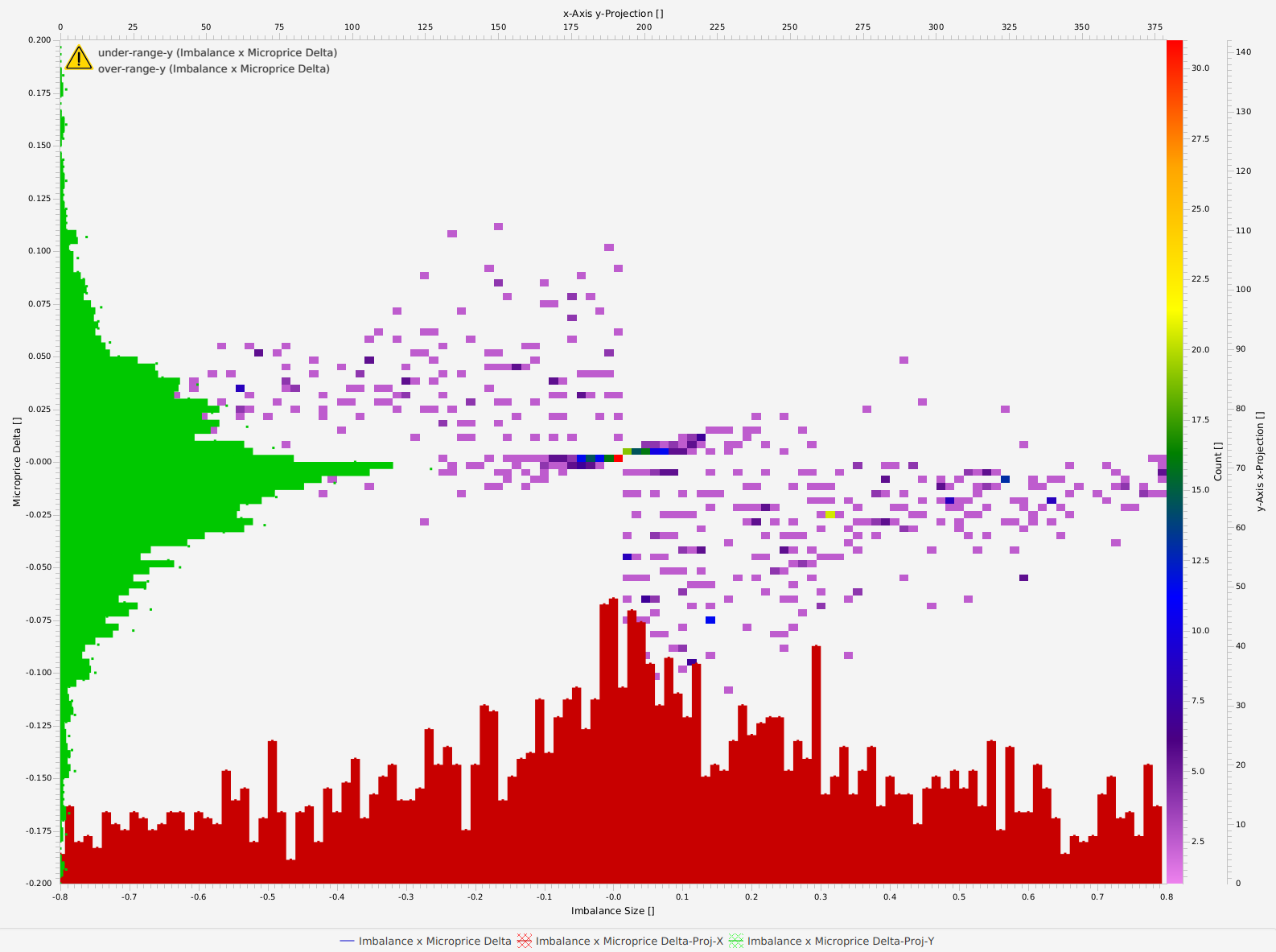}
\caption{Imbalance vs Microprice Adjustment on TSLA on afternoon trading hours}
\label{ref:imbalanceMicroprice2}
\end{figure}

We notice that when the imbalance near zero, namely when the volume on the bid equals the volume on the ask, then the microprice delta is scattered fairly evenly, so there is no immediate bias. However, during times when there is negative skew to the ask side, the microprice deltas tend to be more positive, demonstrating tendency for the latest price to be slightly larger than the current microprice estimate, meaning in general the price will adjust lower.  The same correlation can be found when the imbalance is positively skewed, namely when the volume on the bid is larger than the ask, showing higher demand. With higher demand the latest price is lower than the microprice adjustment in general, showing a clustering of negative values.  

\begin{figure}
\centering
\includegraphics[width=.7\textwidth]{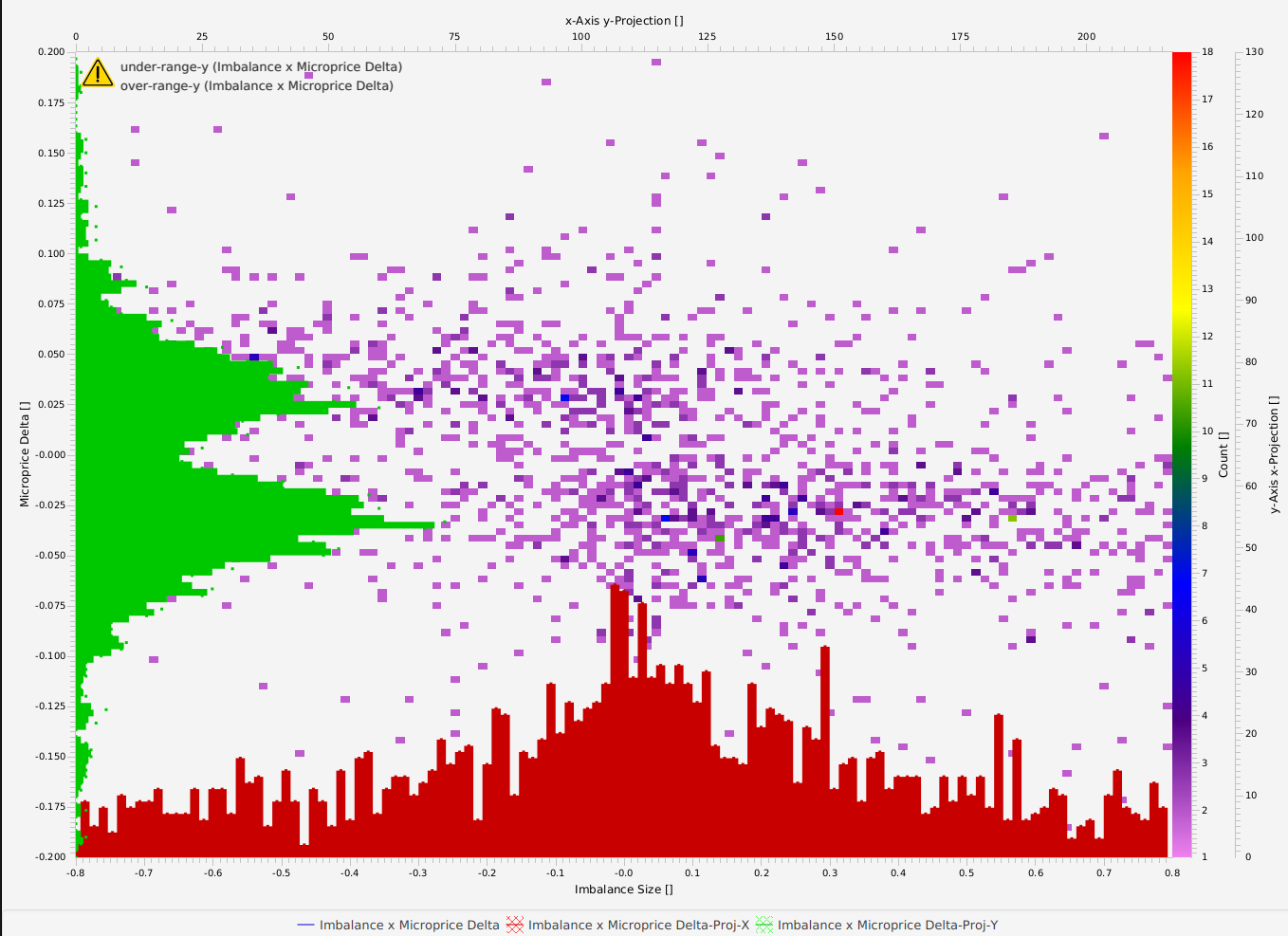}
\caption{Imbalance vs Original Microprice TSLA}
\label{fig:imbalanceMicroprice2}
\end{figure}

When the distribution of latest price minus microprice adjustment centers heavily around zero, for TEM, the imbalance is still quite varied implying there are quite some instances when the microprice adjustment is influence by perhaps higher orderbook imbalances.  In figure \ref{fig:imbalanceMicroprice2} we see that when the imbalance is heavily skewed on the positive side, with bid volume larger than the ask volume (higher demand), then then the latest price minus the microprice adjustment can still have minimal distance. This is not true however when imbalance is skewed to the ask side.

In comparing this with the original microprice estimation shown in \ref{fig:imbalanceMicroprice2}, we see that the histogram on the price difference tends to have a slight skewness on a positive price difference with negative imbalance (skewed on the ask) and negative price difference with positive imbalance. In contradiction to the adjusted microprice estimate, the mean is not zero and the positive or negative adjustments tend are more scattered near the two means of the positive and negative price difference. 

In summary, the microprice adjustments tend to be centered around zero with a much clearer bias depending on the imbalance, demonstrating that the imbalance plays a much more key role in the microprice adjustment estimation.

We now summarize results for the blue chip TSLA and the small cap TEM to compare the average adjusted microprice error and the standard microprice estimate over 6 trading days from the beginning of September 2024. For the errors, we use the $L2$ error from \ref{ref:error}. We also show the average spread size which is presented in dollars, and the estimated volatility which is expressed in percent. In the adjusted error, we express the mean and the standard deviation (in parenthesis) over 10 different TM configurations that were chose randomly. 

\begin{table}[h!]
\centering
\begin{tabular}{|c|c|c|c|}
\hline
\textbf{Adjusted error} & \textbf{Microprice error} & \textbf{AvgSpread} & \textbf{Volatility} \\ 
\hline
0.0619 (0.0149) & 0.0925 & 5.8305 & 56.9 \\ 
0.0635 (0.008) & 0.0773 & 5.7837 & 29.4 \\ 
0.0471 (0.0121) & 0.0745 & 4.9409 & 18.01 \\ 
0.0456 (0.0042) & 0.0583 & 3.6587 & 12.10 \\ 
0.0830 (0.0101) & 0.1002 & 5.0434 & 22.50 \\ 
0.0326 (0.0091) & 0.0375 & 3.1459 & 4.60 \\ 
\hline
\end{tabular}
\vspace{0.5em} 
\caption{Table showing Adjusted, MicroDelta, AvgSpread, and estimated volatility percent values for TSLA}
\label{tab:microdelta1}
\end{table}

We can see that the adjusted microprice accounts for roughly a 10-20 percent improvement and seems to be largely influenced by the average spread size and the volatility for the day. Higher intraday price swings seem to be correlated with a larger microprice adjustment. This could mean that the higher price rank changes in the orderbook seem to be influencing the adjustment during higher price fluctuation.  

\begin{table}[h!]
\centering
\begin{tabular}{|c|c|c|c|}
\hline
\textbf{Adjusted} & \textbf{MicroDelta} & \textbf{AvgSpread} & \textbf{Volatility} \\ 
\hline
0.1630 (0.0202) & 0.1723 & 15.2010 & 10.21 \\ 
0.1099 (0.0155) & 0.1339 & 17.3512 & 31.04 \\ 
0.0676 (0.0332) & 0.1069 & 10.1725 & 11.10 \\ 
0.0648 (0.0092) & 0.0830 & 7.4088 & 35.20 \\ 
0.1173 (0.0206) & 0.1417 & 10.0867 & 25.03 \\ 
0.0461 (0.0091) & 0.0530 & 11.2936 & 9.22 \\ 
\hline
\end{tabular}
\vspace{0.5em} 
\caption{Table showing Adjusted, MicroDelta, AvgSpread, and Volatility values for TEM}
\label{tab:microdelta2}
\end{table}

In table \ref{tab:microdelta2} we apply the same procedure for the small cap TEM over the same 6 trading days using the same TM model parameter configurations. Notice how the average spread sizes are significantly bigger than TSLA, and the estimated volatility tends to be higher.  With the higher volatility and lack of liquidity comes larger variance in estimation of the adjusted microprice. This could mean that the volume changes in the higher bid and ask prices could be more attributed to market spoofing where large orders are readily withdrawn without ever having the intention of matching at that given price.  Given the standard deviation, we see that the microprice adjustment on these small cap microprice estimates seem to not be as effective as in the blue chip microprice estimates.  A more comprehensive study would need to be built in order to determine if this holds across many different blue chips and small caps.

\section{Conclusion}

We demonstrated in this paper an adjustment that can be computed for the microprice that takes into account more information from the orderbook. From the empirical results, we see that including higher bid and ask information from the orderbook along with providing the current microprice and latest imbalance information on the volume can have a positive impact on estimating future prices given the latest matching price. The empirical results so far have demonstrated slightly better accuracy at measuring a microprice adjustment when the spreads are tighter and there the imbalance is centered around zero which is the case often with blue chip equities such as TSLA. For small cap, with larger spread sizes and less matched liquidity at the top bid and ask prices, we see that the microprice adjustments tend to not be  

Of course, in order construct a microprice adjustment in practice on market limit orderbook data in realtime we would clearly need to ensure the efficiency in computation time to gain any proposed benefit. For this we give a quick summary of the computational steps from the adjustment algorithm in \ref{adjustementAlgo}

One way to improve on computational speed is to exploit similarities between consecutive inputs to reuse the previously encoded vector, thereby reducing the required hardware to generate each dimension of the encoded vector. This can have massive impact when the orderbook does not change between states. Storing encoded orderbook vectors for fast retrieval can also play a role in speeding up the input vector for the TM.  An area of future research is in an FPGA implementation exploiting computation reuse which could have a big impact on generating the vectors for the input TM. With a $C$ class TM also having an FPGA implementation as proposed in \cite{prescott2023fpgaarchitectureonlinelearning}, one could have the entire stack run on an FPGA generated from simple orderbook updates. 

\appendix
\section{Tsetlin Machine Architecture}\label{ref:TMs}

\begin{figure}[!ht]
\centering
\includegraphics[width=.7\textwidth]{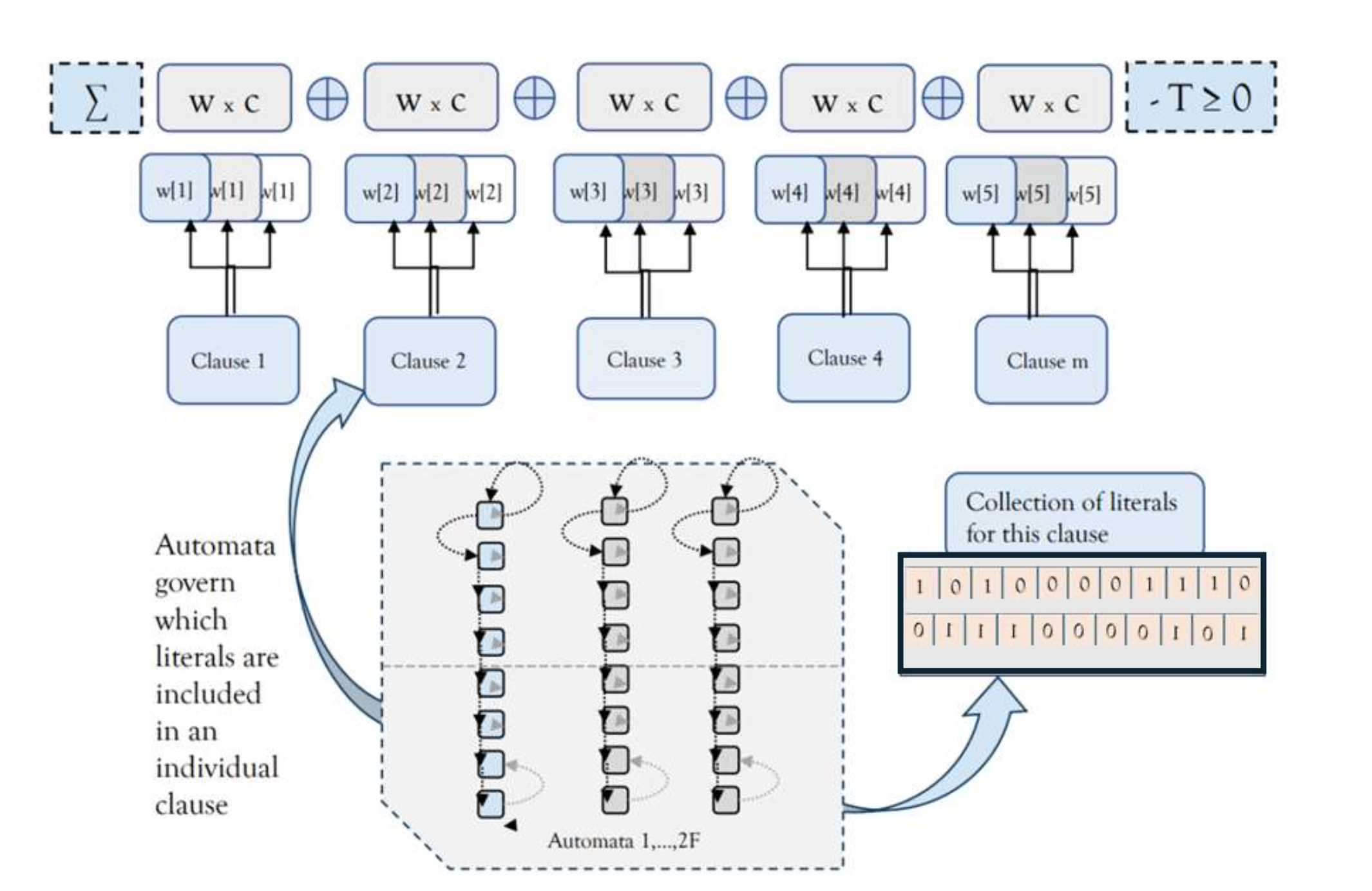}
\caption{Showing a collection of clauses and their relationship with weights that are learned during TM training. Each clause contains a set of literals that are also learned during feedback.}
\label{ref:clause-weights}
\end{figure}

In this paper we employ a recently developed architecture for TM learning originally proposed in \cite{glimsdal2021coalescedmultioutputtsetlinmachines}, a multiclass adaptation of \cite{granmo2018tsetlin}. The architecture shares an entire pool of clauses between all output classes, while introducing a system of weights for each class type. The learning of weights is based on increasing the weight of clauses that receive a Type Ia feedback (due to true positive output) and decreasing the weight of clauses that receive a Type II feedback (due to false positive output). This architectural design allows to determine which clauses are inaccurate and thus must team up to obtain high accuracy as a team (low weight clauses), and which clauses are sufficiently accurate to operate more independently (high weight clauses). The weight updating procedure is given in more detail in \cite{glimsdal2021coalescedmultioutputtsetlinmachines}. Here we illustrate the overall scheme in Figure \ref{ref:clause-weights}. Notice that each clause in the shared pool is related to each output by using a weight dependent on the output class and the clause. The weights that are learned during the TM learning steps and are multiplied by the output of the clause for a given input. Thus clause outputs are related to a set of weights for each output class. 

An additional parameter we use for our TM architecture is to ensure that a maximum number of literals is allowed for every clause. This has the effect of ensuring that the clauses do not become too "dense", proposed in \cite{ostby2024sparsetsetlinmachinesparse}. We will use this architecture for learning encoded features from orderbooks, where the target will be future microprice adjustments. Before we get to that, we introduce the encoding scheme.

\section{Encoding Orderbook features into Sparse Binary Hypervector}\label{ref:hypervector}

Let $\mathbf{v}$ represent a sparse binary vector of length $N$, divided into $S$ segments, where each segment has length $L$. The total length of the vector is thus $N = S \times L$.

Each segment can be denoted as:
\[
\mathbf{v} = [\mathbf{v}_1, \mathbf{v}_2, \dots, \mathbf{v}_S]
\]
where $\mathbf{v}_s$ represents the $s$-th segment of length $L$:
\[
\mathbf{v}_s = (v_{s,1}, v_{s,2}, \dots, v_{s,L}) \quad \text{with} \quad v_{s,j} \in \{0,1\}
\]

A sparse binary vector is characterized by having one bit set to $1$ per segment. In other words, for each segment $\mathbf{v}_s$, there exists one index $k \in \{1, 2, \dots, L\}$ such that $v_{s,k} = 1$, while all other bits are $0$:
\[
\sum_{j=1}^{L} v_{s,j} \leq 1 \quad \forall s \in \{1, 2, \dots, S\}
\]
HV computing is defined by operations for manipulating the vectors in the high-dimensional binary space \cite{Kleyko_2022}. We use two key operations:
\begin{itemize}
    \item Bind (denoted by \( \otimes \)): A vector multiplication-like operation that binds two vectors.
    \item Bundle (denoted by \( \oplus \)): A summation-like operation that bundles multiple vectors together
\end{itemize}

Furthermore, we also include a permutation operation shifts the segments of a sparse binary vector circularly. Given a sparse binary vector $\mathbf{v}$ divided into $S$ segments, the permutation of the vector by $j$ positions is denoted as:
\[
\Pi_j(\mathbf{v}) = [\mathbf{v}_{S-j+1}, \mathbf{v}_{S-j+2}, \dots, \mathbf{v}_S, \mathbf{v}_1, \dots, \mathbf{v}_{S-j}]
\]
where $j$ is the number of positions the segments are shifted. This operation preserves the structure of the sparse vector but changes the order of the segments.

The bind operation, denoted as $\otimes$, combines two sparse binary vectors by performing a circular bitwise shift of one vector's segments based on the indices of the set bits in the other vector's segments. Let $\mathbf{v}$ and $\mathbf{w}$ be two sparse binary vectors with $S$ segments, each of length $L$. The bind operation is defined segment-wise as:
\[
(\mathbf{v} \otimes \mathbf{w})_s = \text{rotate-right}(\mathbf{w}_s, \text{index of } 1 \text{ in } \mathbf{v}_s)
\]
where $\text{rotate-right}(\mathbf{w}_s, k)$ denotes a circular right shift of the segment $\mathbf{w}_s$ by $k$ positions, with $k$ being the index of the bit set to $1$ in the corresponding segment of $\mathbf{v}_s$.

The bundle operation aggregates multiple sparse binary vectors into a single sparse binary vector based on majority voting. Let $\mathbf{v}^{(1)}, \mathbf{v}^{(2)}, \dots, \mathbf{v}^{(n)}$ represent $n$ sparse binary vectors. For each segment $s$, the bundle operation is defined as follows:
\[
\mathbf{b}_s[j] = \begin{cases}
1 & \text{if the majority of vectors have a $1$ at index } j \\
0 & \text{otherwise}
\end{cases}
\]
If there is a tie between multiple indices in a segment, one index is randomly selected. The final bundled vector is:
\[
\mathbf{b} = [\mathbf{b}_1, \mathbf{b}_2, \dots, \mathbf{b}_S]
\]

\bibliographystyle{unsrt}  
\bibliography{references}

\end{document}